\documentclass[11pt,twoside]{article}  
\usepackage{asp2006}
\usepackage{adassconf}

\begin{document}   

\paperID{E.48}

\title{The NOAO NEWFIRM Pipeline}
       
\markboth{Swaters and Valdes}{The NOAO NEWFIRM Pipeline}

\author{Robert A. Swaters}
\affil{Department of Astronomy, University of Maryland, MD, USA}

\author{Francisco Valdes, Mark E. Dickinson}
\affil{National Optical Astronomy Observatory, Tucson, AZ, USA}

\contact{Rob Swaters}
\email{swaters@astro.umd.edu}

\paindex{Swaters, R.~A.}
\aindex{Valdes, F.}


\setcounter{footnote}{1}

\begin{abstract}
The NOAO NEWFIRM Pipeline produces instrumentally calibrated data
products and data quality measurements from all exposures taken with
the NOAO Extremely Wide-Field Infrared Imager (NEWFIRM) at the KPNO
Mayall 4-meter telescope. We describe the distributed nature of the
NEWFIRM Pipeline, the calibration data that are applied, the data
quality metadata that are derived, and the data products that are
delivered by the NEWFIRM Pipeline.
\end{abstract}

\section{Introduction}

The NOAO Extremely Wide-Field Infrared Imager (NEWFIRM) is a 1 to 2.4
micron IR camera for both the Kitt Peak and CTIO 4m telescopes. The
instrument is equipped with broad and narrow band filters. The focal
plane of NEWFIRM consists of four 2048 by 2048 detectors laid out in a
2 by 2 mosaic. The field of view is 27.6' by 27.6' (including a 35''
wide gap between the detectors). The pixel size is 0.4''.

The data taken with NEWFIRM are processed by the NOAO NEWFIRM
pipeline. This pipeline consists of two distinct components. The first
is the Quick Reduce Pipeline (QRP), which produces reduced data in
near-real time at the telescope. The second component is the Science
Pipeline (SP). It produces uniformly reduced, high-quality data at the
end of an observing block. The products of the SP will be made
available to the observer and through the NOAO archive.

Both QRP and SP apply all basic calibration steps, such as dark
subtraction, flat fielding, and sky subtraction. WCS solutions are
also determined for each exposure by matching stars against the 2MASS
catalog (Skrutskie et al.,\ 2006). The stars in common with 2MASS are
also used to determine a first-order photometric calibration.  The
QRP, as its final data product, combines data for the same pointing to
create deep dither stacks. The SP continues, and uses these first
stacks to create object masks, which are then used to mask out objects
in the individual exposures for an improved second-pass sky
subtraction and deep dither stacks. In addition, the SP also deals
with the persistence seen in the NIR detector, large-field mosaicked
images, and data quality metrics for calibration and science data.

\section{A Distributed Pipeline System}

Both the QRP and the SP make use of the NOAO High-Performance Pipeline
System (Scott et al.\ 2007). This system is designed to run on a
cluster of computers, and it allows easy and efficient use of CPU
resources in problems with inherent parallelism such as the processing
of observations from mosaic cameras.

In the pipeline system, the data reduction process has been organized
into a hierarchical structure of different individual pipelines. Each
of these pipelines deals with one aspects of the reduction process
e.g., constructing calibration data (such as darks, dome flats, or
master sky flats), applying these calibration data, collecting the
final data products, and controlling other pipelines. All of these
pipelines, in turn, are subdivided into a set of modules, each of
which carries out a small step of the data reduction process. These
modules can be written in any language. Because the pipeline system
uses IRAF, most modules are IRAF scripts.

\begin{figure}[t]
\epsscale{1.0}
\plotone{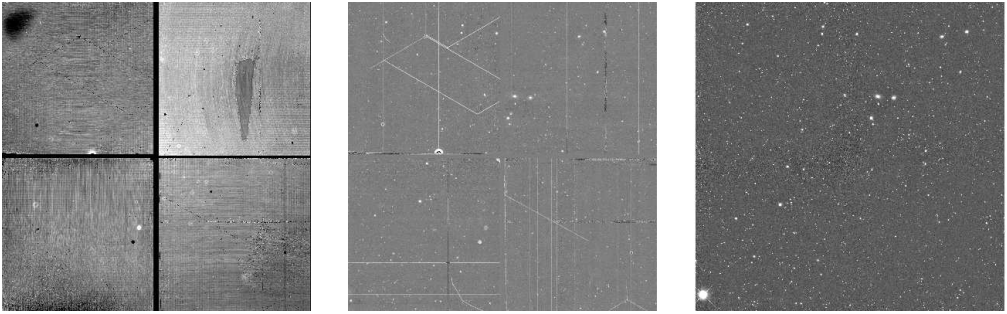}
\caption{An example of sparse field observed with NEWFIRM. The left
  panel shows a raw image, and the middle panel shows a single dark-
  subtracted, flat-fielded, and sky-subtracted image. The right panel
  shows a fully reduced stack of multiple
  observations.} \label{O4.1-fig-1}
\end{figure}

All of the pipelines and associated modules can be configured to run
on all available nodes (or any subset thereof), ensuring maximum use
of available resources.  Whenever a pipeline is to be started, a node
selection algorithm is called to find the best node by considering
issues such as load and minimizing network traffic.  Distribution of
the work across the nodes in the cluster also depends on the nature of
the steps in the data reduction process, and processing may be
distributed for example by detector or by multi-extension fits (MEF)
files.  An example of distribution by detector is the calculation of
the master sky flats from a group of observations. In this case, the
data from the 4 different NEWFIRM detectors are sent to 4 different
nodes, each of which will process the data from one unique array. When
applying calibration data to MEF files, however, it is more efficient
to leave the individual arrays on the same node, because this
significantly reduces transfer of data between nodes. In this case,
the data are distributed across nodes by MEF files.

To make sure available resources are used efficiently, any node can
run multiple instances of a pipeline. Thus, on a node with two CPUs,
two instances of a pipeline that apply calibration data to science
observations will run in parallel.

\section{Calibration}

Both the QRP and the SP apply the major calibrations to the
raw data:

\begin{itemize}

\item Apply a correction for the inherently nonlinear nature of the
  NIR detectors
\item Subtract the dark structure (from dark exposures)
\item Apply flat fields
\item Flag any bad and saturated pixels, and pixels affected by
  persistence)
\item Do a first-pass sky subtraction using sky images constructed
  from other images in the observing sequence (for sparse field data)
  or from an offset sky field (e.g., when observing large objects).
\item For object exposures, determine the world coordinate system and
  a rough photometric zeropoint by matching objects in the field
  against the 2MASS catalog
\end{itemize}

For the QRP, the emphasis is on speed of processing while still
producing data that are reduced thoroughly and allow the users to
evaluate the quality and depth of their data.  The SP generates
advanced data products by also including the following time-consuming
steps:

\begin{itemize}
\item 2nd pass sky sky subtraction (masks are used to remove objects
  when exposures are combined to create sky images)
\item Detection and flagging of transient events (e.g., cosmic ray
  hits, satellite and airplane trails)
\item Object masking in offset-sky exposures for improved sky
  subtraction
\end{itemize}

\begin{figure}[t]
\epsscale{0.8}
\plotone{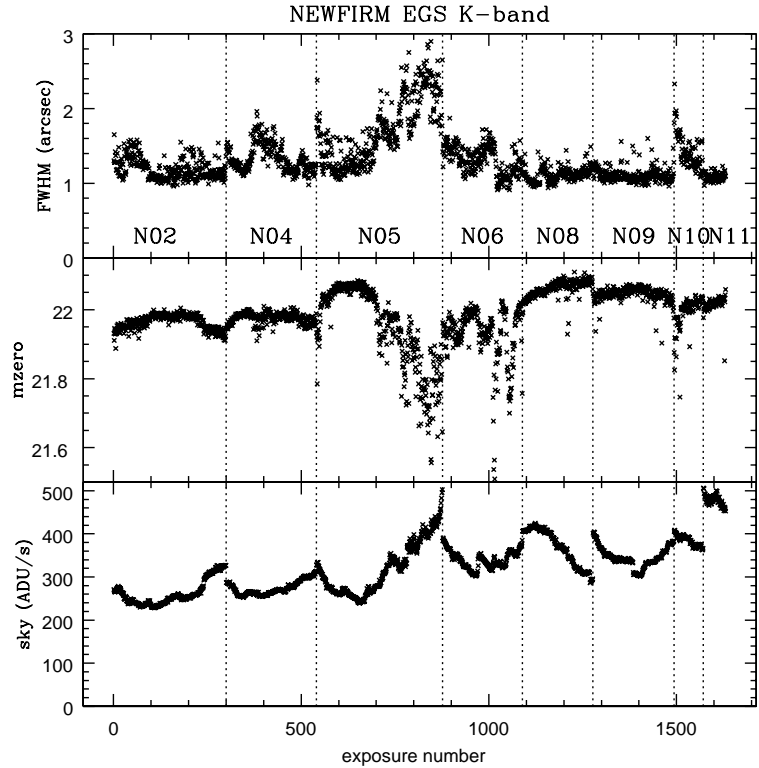}
\caption{An example of the observing conditions as reported by the
  QRP, showing, from top to bottom, the seeing, the photometric
  zeropoint, and the sky level for K-band observations across 11
  nights.} \label{O4.1-fig-1}
\end{figure}

\section{Data Quality}

The QRP and the SP verify, check, and characterize the data
being processed throughout the reduction process, and the resulting
metadata is stored. This is done for several reasons:

\begin{itemize}
\item Immediately after ingesting the data, the fits files and headers
  are verified and data that cannot be processed are
  rejected. Examples are: corrupt or incomplete fits files, fits files
  with missing critical header keywords. When possible, missing
  keywords are reconstructed.
\item Statistics and other numerical characterizations of individual
  calibration exposures are compared against expected values and
  outliers are rejected.
\item Critical metrics that help users evaluate the quality of the
  data, such as seeing, photometric depth, sky level, and WCS are
  recorded. In the case of the QRP, these metrics are presented to the
  user both for the most recent data, and also for prior observations
  for trending purposes.
\end{itemize}

For the SP, a more exhaustive set of metrics is measured
for calibration and science data. These metadata, derived from the raw
data, intermediate steps, and the final pipeline data products are all
stored in the pipeline metadata database. This database can be queried
by the pipeline itself to carry information from one module to the
next, but it also provides a means to investigate long term trends in
the instrument's performance.

\section{Data products}

After the exposures have been fully calibrated, the data are
reprojected to the same orientation and pixel size, and to the closest
tangent point on a predefined grid (thus ensuring that spatially close
exposures have the same tangent point), these exposures are then
combined into one stacked image, which could span a large field of
view, go deep, or both.  The original reduced image, the resampled
one, and the stacked one are end products of the pipeline.  In
addition, the pipeline produces masks for each of these.

\end{document}